\documentclass[twocolumn,aps,pre,showpacs,floatfix]{revtex4}
\usepackage{graphicx}
\usepackage{amssymb}
\usepackage{amsmath}
\usepackage{bm}

\begin{document}

\title{Coherent periodic activity in excitatory 
Erd\"os-Renyi neural networks:\\ 
The role of network connectivity}


\author{Lorenzo Tattini}
\email{lorenzotattini@gmail.com}    
\affiliation{CNR - Consiglio Nazionale delle Ricerche -
Istituto dei Sistemi Complessi, via Madonna del Piano 10,
I-50019 Sesto Fiorentino, Italy}

\author{Simona Olmi}
\email{simona.olmi@fi.isc.cnr.it}    
\affiliation{CNR - Consiglio Nazionale delle Ricerche -
Istituto dei Sistemi Complessi, via Madonna del Piano 10,
I-50019 Sesto Fiorentino, Italy}
\affiliation{INFN - Sezione di Firenze and CSDC, via Sansone 1, 50019 Sesto      Fiorentino, Italy}

\author{Alessandro Torcini}
\email{alessandro.torcini@cnr.it}    
\affiliation{CNR - Consiglio Nazionale delle Ricerche -
Istituto dei Sistemi Complessi, via Madonna del Piano 10,
I-50019 Sesto Fiorentino, Italy}
\affiliation{INFN - Sezione di Firenze and CSDC, via Sansone 1, 50019 Sesto      Fiorentino, Italy}

\begin{abstract}
In this article we investigate the role of connectivity in
promoting coherent activity in excitatory neural networks.
In particular, we would like to understand if the onset of collective
oscillations can be related to a minimal average connectivity 
and how this critical connectivity depends on the number of
neurons in the networks.
For these purpouses, we consider an excitatory random network of leaky 
integrate-and-fire pulse coupled neurons.
The neurons are connected as in a directed
Erd\"os-Renyi graph with average connectivity $\langle k \rangle$ scaling
as a power law with the number of neurons in the network.
The scaling is controlled by a parameter $\gamma$,
which allows to pass from massively connected to sparse networks
and therefore to modify the topology of the system.
At a macroscopic level we observe two distinct dynamical
phases: an Asynchronous State (AS) corresponding to a 
desynchronized dynamics of the neurons and a Partial Synchronization (PS) 
regime associated with a coherent periodic activity of the network. 
At low connectivity the system is in an Asynchronous State, while PS emerges 
above a certain critical average 
connectivity $\langle k \rangle_c$. For sufficiently large networks,  
$\langle k \rangle_c$ saturates to a constant 
value suggesting that a minimal average connectivity is sufficient to observe 
coherent activity in systems of any size irrespectively of the kind of considered network:
sparse or massively connected. However, this value depends on the nature of the synapses:
reliable or unreliable. For unreliable synapses the critical 
value required to observe the onset of macroscopic behaviors is noticeably smaller than 
for reliable synaptic transmission. Due to the disorder present in the 
system, for finite number of neurons we have inhomogeneities
in the neuronal behaviors, inducing a weak form of chaos,
which vanishes in the thermodynamic limit.
In such a limit the disordered systems exhibit regular (non chaotic) dynamics 
and their properties correspond to that of a homogeneous fully connected network 
for any $\gamma$-value. Apart for the peculiar exception of sparse networks,
which remain intrinsically inhomogeneous at any system size.
\end{abstract}

\keywords{Coherent activity, Random neural networks, Integrate-and-Fire Neurons,
Erd\"os-Renyi Graph, Reliable and unreliable synapses, Lyapunov analysis}

\pacs{05.45.Xt, 84.35.+i, 87.19.lj, 87.19.ln}


\maketitle
{\bf

The spontaneous emergence of collective dynamical
behaviours in random networks made of many (identical) interacting units 
is a subject of interest in many different research fields 
ranging from biological oscillators to power grids.
In particular, how the macroscopic dynamics of the network
is influenced by the topology is an active research line
not only for nonlinear dynamics, but also
for many other scientific disciplines, as for example
(computational) neuroscience~\cite{grinstein}.
However, the most part of the performed analysis have been 
devoted to the emergence of the fully synchronized regimes,
but in neuroscience a complete synchronization is usually
a symptom of neural disorders, while coherent oscillations
are often associated to a partial synchronization among neurons
during brain activity~\cite{buszaki}.
Coherent oscillatory activities are prominent in the cortex of the 
awake brain during attention, and have been implicated in higher level 
processes, such as sensory binding, storage of memories, and even
consciousness.

In this article, we analyse how the presence of 
disorder in the connections can influence the emergence of 
coherent periodic activity in Erd\"os-Renyi networks of excitatory pulse coupled 
spiking neurons. Our main result indicates that the parameter controlling the transition 
from asynchronous to coherent neural activity is simply the average connectivity.
Furthermore, for (sufficiently large) networks
the critical value of the average connectivity turns out to be independent
of the network realization, sparse or massively connected,
but it is instead influenced by the nature of the disorder, quenched or annealed.
}

\section{Introduction}
\label{intro}
Neural collective oscillations have been observed in very many 
context in brain circuits, ranging from ubiquitous $\gamma$ oscillations
to $\theta$ rhythm in the hippocampus.  The origin of these oscillations 
is commonly associated with the
balance between excitation and inhibition in the network,
while purely excitatory circuits are believed to lead 
to "unstructured population bursts" \cite{buszaki}. 
However, recent `ex vivo'' measurements performed on the
rodent neocortex~\cite{allene} and hippocampus~\cite{bonifazi}
in the early stage of brain maturation reveal 
coherent activity patterns, such as {\it Giant Depolarizing Potentials}.
These collective oscillations emerge 
despite the fact that the GABA transmitter has essentially
an excitatory effect on immature neurons~\cite{review}.
Therefore, also in purely excitatory networks 
one can expect non trivial dynamics at a macroscopic level.

Numerical and theoretical studies of collective motions
in networks of simple spiking neurons have been mainly devoted to 
balanced excitatory-inhibitory configurations, e.g see
\cite{brunel2000} and references therein. Only few studies
focused on coherent periodic activity in fully coupled
excitatory networks of leaky integrate-and-fire (LIF) neurons.
These analysis revealed a regime characterized by a {\it partial synchronization} (PS)
at the population level, while the single neurons perform quasi-periodic motions~\cite{vvres}.
It has been shown that the PS regime is quite robust to perturbations, 
since it survives to moderate levels of noise or dilution~\cite{mohanty,olmi}.

Furthermore, in their recent study Bonifazi et al. \cite{bonifazi} found
that the functional connectivity of developing hippocampal networks 
is characterized by a truncated power-law distribution of the out-degrees
with exponent $\gamma = 1.1 - 1.3$. This scaling has been shown to hold over
one/two decades, thus not ensuring a scale-free distribution for the links over
all scales, but surely indicating the presence of a large number of hub neurons, 
namely cells characterized by a high connectivity. 
At early developmental stages of the brain,
GABAergic hub interneurons, performing complex excitatory/shunting inibitory 
actions~\cite{review}, seem to be responsible for the 
orchestration of the coherent activity of hippocampal networks~\cite{bonifazi}.
The relevance of hubs in rendering a neural circuit extremely hyperexcitable has 
been also demonstrated in simulation studies of a realistic model of the epileptic rat 
dentate gyrus, even in the absence of a scale-free topology \cite{morgan}.

Motivated by these studies, but without attempting to reproduce the experimental 
results, we focus on a very preliminary question: to which extent is the macroscopic neural dynamics
influenced by the average degree of connectivity of the neurons ?
Our specific aim is to analyze the key ingredients leading to the 
onset of coherent activity, as opposed to asynchronous dynamics, for different 
network size and topology.

Specifically, we consider the transition from an asynchronous 
regime to partial synchronization in the excitatory LIF pulse coupled neural 
networks introduced by Abbott and van Vreeswijk~\cite{abbott}.
At variance with previous works, we consider dynamical evolution on 
random Erd\"os-Renyi (ER) networks with an average connectivity corresponding to that
of a truncated power-law distribution with a decay exponent $1 <  \gamma <  2$.
This amounts to have an average connectivity which scales proportionally
to $N^{2-\gamma}$ with the number of neurons in the network.
In the limit $\gamma \to 1$ the massively 
connected network, where the connectivity is proportional to $N$, is recovered;
while for $\gamma \to 2$ a sparse network, where
the average probability to have a link between two neurons
vanishes in the thermodynamic limit \cite{golomb}, is retrieved.
The topology of ER networks is modified by varying the parameter $\gamma$ in 
the interval $[1:2]$, in particular as far as $\gamma \le 2$ trees and cycles of any order
are present in the network, while for $\gamma \to 1$ complete subgraphs
of increasing order appear in the system \cite{reti}.
In the paper we will study how and to which extent these topological modifications influence
the macroscopic dynamics of the network, with particular emphasis on the transition
from asynchronous to (partially) synchronous collective dynamics.

The paper is organized as follows:
the next Section is devoted to the introduction of the neural 
model and of the indicators employed to characterize the dynamics
of the network. The phase diagram reporting the collective
states emerging in our system is described in Sect. 3. 
The influence of finite size effects on the coherent activity is
analyzed in Sect. 4, while a characterization of the neural dynamics
in term of maximal Lyapunov exponent is reported in Sect. 5. 
A brief discussion of our results is outlined in Sect. 6.

\section{Model and Methods}
\label{sec:1}

\subsection{The Model} 
We study a network of $N$ LIF neurons with the membrane 
potential $x_i(t) \in [0:1]$ of the neuron $i$ evolving as:
\begin{equation}\label{eq:x1}
\dot{x}_{i}(t)= a-x_{i}(t)+I_i(t)\, \quad\quad i=1,\cdots, N
\\ \quad ,
\end{equation}
where $a > 1$ is the suprathreshold DC current, $I_i$ the synaptic
current. Whenever the neuron reaches the threshold $x_i=1$, a
pulse $s(t)$ is instantaneously transmitted to all the connected
post-synaptic neurons and the membrane potential of neuron $i$ is
reset to $x_i=0$. The synaptic current can be written as
$I_i(t) = g E_i(t)$, with $g  >  0$ representing
the synaptic excitatory strength while the field $E_i(t)$ is 
the linear superposition of the pulses $s(t)$ received 
by neuron $i$ in the past, in formula
\begin{equation}\label{eq:E0}
E_i(t)= \frac{1}{k_i} \sum_{n|t_n < t} C_{j,i} 
\Theta(t-t_n) s(t-t_n) \quad ,
\end{equation}
where $k_i$ is the number of pre-synaptic neurons connected to the
neuron $i$ (i.e. the in-degree of neuron $i$) and $\Theta(t)$ the causal $\Theta$ function.
The connectivity matrix $C_{j,i}$ appearing in Eq. (\ref{eq:E0})
has entry $1$ (resp. 0) depending if the pre-synaptic neuron $j$ 
is connected (resp. not connected) to neuron $i$ and in general
it is not symmetric.

Following van Vreeswijk ~\cite{vvres} we assume, for the single pulse emitted 
at $t=0$, the shape $s(t)=\alpha^2 t \exp(-\alpha t)$.
The explicit equation (\ref{eq:E0}) can be thus rewritten as an
implicit ordinary differential equation:
\begin{equation}\label{eq:E}
\ddot E_i(t) +2\alpha\dot E_i(t)+\alpha^2 E_i(t)= 
\frac{\alpha^2}{k_i}\sum_{n|t_n\langle t} C_{j,i} \delta(t-t_n) \ .
\end{equation}

The continuous time evolution of the network can be transformed
in a discrete time event-driven map 
by integrating Eq. ~(\ref{eq:E}) from time $t_n$ to time $t_{n+1}$,
$t_n$ being the time immediately after the $n$-th spike emission.

Following Olmi et al. \cite{olmi}, the event-driven map read as:
\begin{eqnarray}\label{eq:map}
E_i(n+1)&=&E_i(n) {\rm e}^{-\alpha \tau(n)}+Q_i(n)\tau(n) 
{\rm e}^{-\alpha \tau(n)} \\
\label{qq}
Q_i(n+1)&=&Q_i(n)e^{-\alpha \tau(n)}+C_{j,i}\frac{\alpha^2}{k_i} \\
x_{i}(n+1)&=&x_i(n)e^{-\tau(n)}+a(1-e^{-\tau(n)})+g H_i(n) \, .
\end{eqnarray}
where $Q_i \equiv \alpha E_i+\dot E_i$ is an auxilary
variable, $H_i=H_i(E_i,Q_i,\tau)$ is a 
nonlinear function and
$\tau(n)= t_{n+1}-t_n$ is the interspike time interval. 
This can be determined by solving the following implicit relationship
\begin{equation}\label{eq:ti2}
\tau(n)=\ln\left[\frac{a-x_m(n)}{a+gH_m(n)-1}\right] \ ,
\end{equation}
where $m(n)$ identifies the neuron which will fire next 
at time $t_{n+1}$ by reaching the threshold value $x_m = 1$.

The evolution of the system is now modeled with a discrete time map of $3 N - 1$
variables, $\{E_i,Q_i,x_i\}$. In fact the degree of freedom associated
with the membrane potential of the firing neuron has been
removed by the implementation of the Poincar\'e-section, leading to the event
driven map. More details on the model are reported in Ref. ~\cite{olmi};
however at variance with that study the pulse amplitudes, appearing
in eqs. (\ref{eq:E0}),(\ref{eq:E}) and (\ref{qq}), are normalized
by the in-degree $k_i$ of neuron $i$ and not by the total number of 
neurons $N$.

The model parameters were fixed as $a=1.3$, $g=0.4$ and $\alpha=9$,
in order to ensure the emergence of a PS regime 
in the corresponding fully coupled network ~\cite{vvres}. 

\subsection{The Connectivity Matrix} 
Our choice for the connectivity matrix has been mainly motivated by
the results obtained for ``ex vivo'' cells by Bonifazi et al. \cite{bonifazi}.
In particular, these authors have shown that in developing hippocampal networks the functional connectivity
is distributed according to power-law distribution, at least over one/two
decades.  The distribution of links $k$ per neuron is thus given by $P(k) = p k^{-\gamma}$,
where $p$ is a normalization constant. Therefore, for a finite network made of
$N$ neurons the average connectivity for the truncated (power-law) distribution is:
\begin{equation}
\langle k \rangle  =  \frac{p}{2-\gamma} \left[ N^{2-\gamma} -1 \right]
\quad .
\label{kave}
\end{equation}
Furthermore, Bonifazi et al \cite{bonifazi} have measured quite low
values for the exponent $\gamma$, namely  $\gamma= 1.1 - 1.3$,
suggesting the existence of a large number of 
highly connected neurons, hubs, in the network. The role of hub neurons 
in orchestrating the level of synchrony in living brain networks
has been clearly demonstrated in \cite{bonifazi}.
In spite of these findings the underlying network topology is maybe less 
important than the actual number of highly connected neurons, 
as suggested by Morgan and Soltesz \cite{morgan}. 

In order to single out
the effect of the average number of connections on the network dynamics 
we decided to limit our analysis to Erd\"os-Renyi (ER) random network~\cite{reti},
but with an average connectivity given by Eq. (\ref{kave}). In particular, we considered
a directed ER random graph, where the distribution of links
is well approximated by a Poisson distribution~\cite{reti}, namely:
\begin{equation}
P(k)  =  {\rm e}^{-\langle k \rangle} \frac{\langle k \rangle^k}{k !} 
\quad .
\label{poisson}
\end{equation}
According to Eq. (\ref{poisson}) the degree distribution is completely defined once
the value of $\langle k \rangle$ is given. As a matter of fact, by choosing
for $\langle k \rangle$ the expression (\ref{kave}), the probability
of existence of an unidirectional link connecting neuron $j$ to $i$
(i.e. the probability to have $C_{j,i}=1$) is:
\begin{equation}
Pr(N,\gamma)  =  \frac{\langle k \rangle}{N} =
\frac{p}{2-\gamma} \left[ N^{1-\gamma} -\frac{1}{N} \right]
\quad .
\label{prob}
\end{equation}
In the limit $\gamma \to 1$ the massively
connected network is recovered~\cite{golomb}, since the average connectivity
$\langle k \rangle = p \times (N-1)$ is proportional to the system size and
$Pr(N,1) = p(1-1/N)$ is, apart finite-size corrections, constant
and coincident with $p$. For $\gamma <  2$ (resp. $\gamma  > 2$) the
average number of synaptic inputs per neuron will grow (resp. decrease) with $N$,
in the limiting case $\gamma =2$ a sparse network will be essentially recovered 
\cite{note1} since $\langle k \rangle = p \ln N$ will vary in a limited manner with respect to the system
size. Indeed, by varying $N$ by three orders of magnitude
from 100 to 100,000 the value of $\langle k \rangle$ will modify
from 3.7 to 9.2 with $p=0.8$.
In the following, we studied networks of various sizes $N$,
ranging from $N=100$ to $N=200,000$, for different $\gamma$-values 
in the interval $[1:2]$. The value $p$ is usually fixed to 0.8,
apart for $\gamma=1$ (constant probability case)
where the dependence on $p$ is examined in details in Section 3.

Let us stress that in the present study the distributions of
pre-synaptic (in-degree) and of post-synaptic (out-degree) connections are identical,
and this is guaranteed by the above outlined procedure to determine 
unidirectional links for ER networks. 

In what follows, we consider two different ways to select
the random connectivity matrix: in the first case the synaptic connections are 
randomly chosen and are constant in time ({\it quenched disorder}); while for the
second procedure the neurons receiving the excitatory pulse are randomly selected
each time a neuron fires ({\it annealed disorder}).
The latter choice can be justified from a physiological point
of view by the fact that the synaptic transmission of signals 
is an {\it unreliable} process \cite{Kinzel2008}.
It should be noticed that in the annealed case, 
since the network modifies in time, 
the pulse amplitudes are  normalized by the average 
in-degree $\langle k \rangle$ and not by $k_i$ as in the quenched case.

\subsection{Characterization of Macroscopic Attractors} In order to perform a
macroscopic characterization of the dynamical states of the network we 
exploit the average fields:
\begin{equation}
\label{eq:avefields}
\bar E(t) = \frac{g}{N}\sum_{i=1}^N E_i(t) \qquad , \qquad
\bar Q(t) = \frac{g}{N} \sum_{i=1}^N Q_i(t)  \ ;
\end{equation}
where the fields have been also rescaled by the synaptic strength
as done in Ref. \cite{olmi}.

As a measure of the level of homogeneity among the neurons 
of the network, we consider the standard deviation $\sigma(t)$ of
the fields $\{E_i(t)\}$ acting on the single neurons
\begin{equation}
\label{eq:sigma}
\sigma(t) = 
\left( \frac{g^2}{N} \sum_{i=1}^{N}E_{i}^{2}({t})-\bar{E}^{2}({t})\right)^{1/2} \, ;
\end{equation}
for completely homogeneous systems, such as globally coupled networks,
$E_i(t) \equiv \bar{E}$ and $\sigma \equiv 0$.

The degree of synchronization among the neurons is quantified by
the order parameter usually employed in the context of phase oscillators~\cite{kurabook}
\begin{equation}
\label{kura}
R(t) =\left| \frac{1}{N} \sum_{j=1}^N {\rm e}^{i \theta_j(t)} \right|  \ ,
\end{equation}
where $\theta_j$ is the phase of the $j^{\footnotesize\textnormal{th}}$ neuron, that can be properly
defined as a (suitably scaled) time variable~\cite{winfree},
$\theta_j(t) = 2\pi (t-t_{j,n})/T_{q,n}$, where $t_{j,n}$
indicates the time of the last spike emitted by the $j^{\footnotesize\textnormal{th}}$ neuron, 
while $T_{q,n}=t_{q,n+1}-t_{q,n}$ is the $n$-th interspike interval associated to
the neuron $q$ which was the last to fire in the network \cite{note2}.

A non zero-value of $R$ represents an indication of {\it partial} synchronization 
among the neurons, while a vanishingly small $R \sim 1/\sqrt{N}$ 
is observable for asynchronous states in finite systems~\cite{kurabook}.

\subsection{Lyapunov Analysis} The dynamical microscopic instabilities
of a system can be characterized in terms of the maximal Lyapunov exponent 
$\lambda$: a positive $\lambda$ being a measure of the degree of chaoticity
of the considered system. In particular, we have employed the standard method
developed in Ref. \cite{lyap} to estimate the maximal Lyapunov exponent
by following the evolution of an infinitesimal perturbation to a 
reference trajectory.

\section{Phase Diagram}
\label{sec:2}

\begin{figure}
\includegraphics[width=0.4\textwidth]{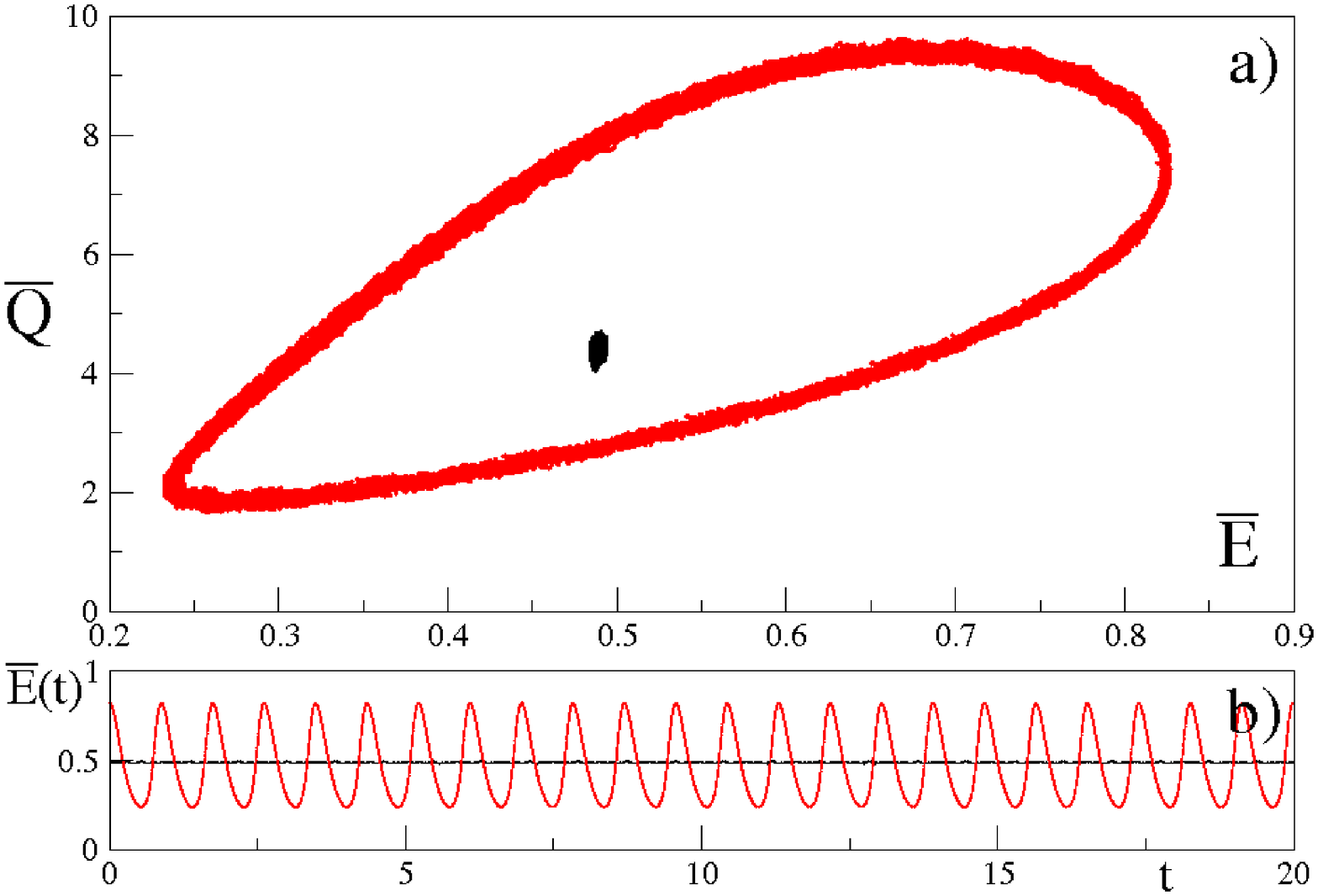}
\includegraphics[width=0.4\textwidth]{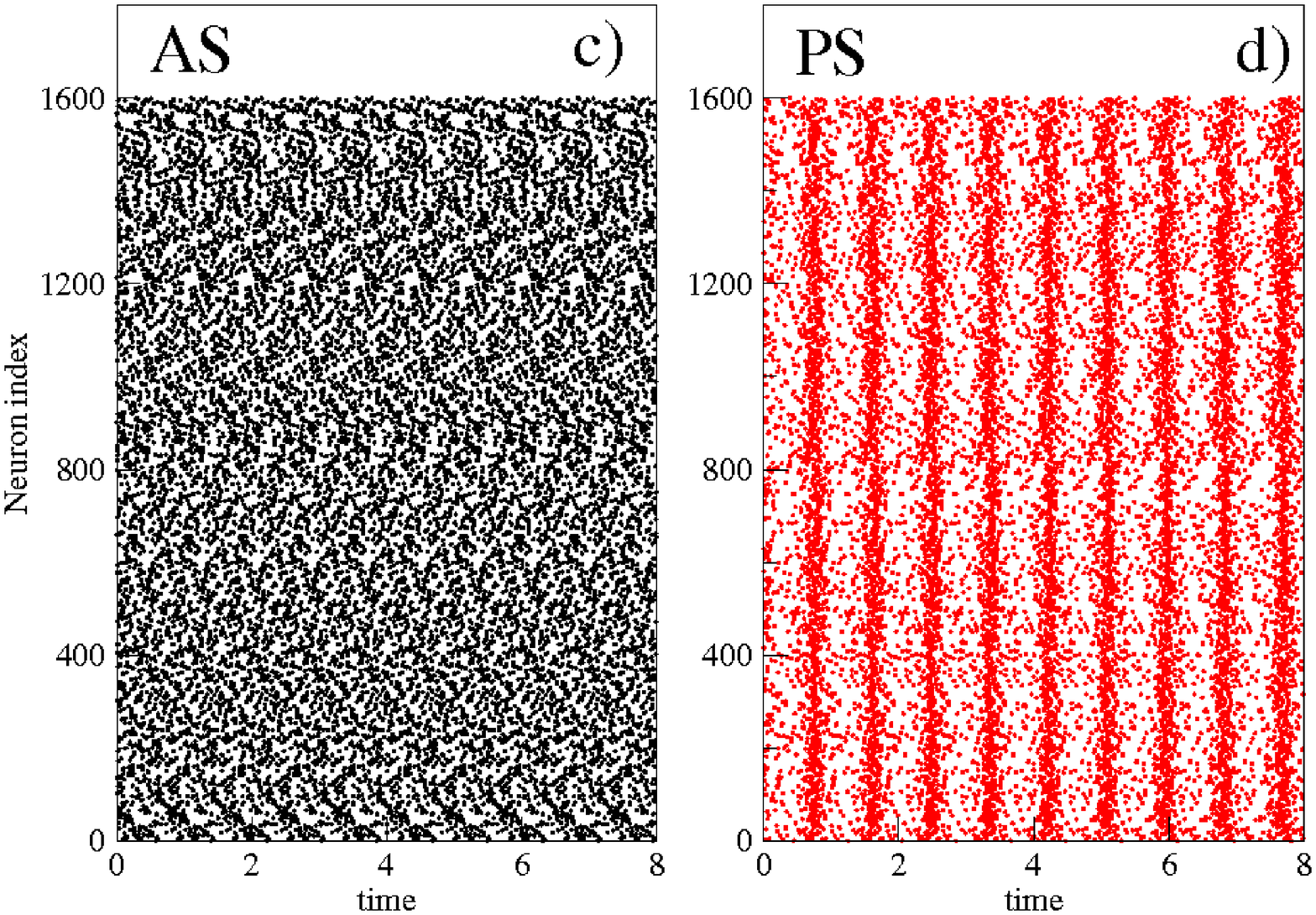}
\caption{(Color online) AS and PS characterization in terms of 
macroscopic fields and single neuron dynamics.
Panel a: macroscopic attractors in the $(\bar E, \bar Q)$ plane.
Panel b: the average field $\bar E$ as a function of time.
Panel c and d: raster plots.
The data refer to ER networks with $\langle k \rangle=p \times N$, quenched
disorder, and $N=1,600$, the black (resp. red) symbols correspond to
AS observable for $p=0.2$ (resp. PS for $p=0.7$).
}
\label{fig:1}       
\end{figure}

In this model two different macroscopic regimes
can be observed: the asynchronous state (AS)
and the partial synchronization (PS).
AS is characterized by an incoherent dynamics
of the neurons in the network leading to a spot-like macroscopic attractor 
in the $(\bar E, \bar Q)$ plane and an almost constant 
average field $\bar E$~\cite{abbott}, while to the coherent PS regime corresponds
a closed curve attractor~\cite{mohanty} and a periodic behavior
of $\bar E$ in time as shown in Fig. \ref{fig:1}a and b. The incoherent
and coherent neural dynamics can be clearly 
appreciated, in the two regimes, also at microscopic level by examining the corresponding 
raster plots reported in Fig.  \ref{fig:1}c and \ref{fig:1}d.

As a first aspect, we will investigate the occurrence of AS and PS
for finite networks and different average connectivity $\langle k \rangle$.  
In order to distinguish the two regimes we have examined the macroscopic attractor
shape, the extrema values of the average field $\bar E$, and the synchronization 
parameter $R$ as a function of $N$.

\subsection{ER networks with constant probability} 

\begin{figure}
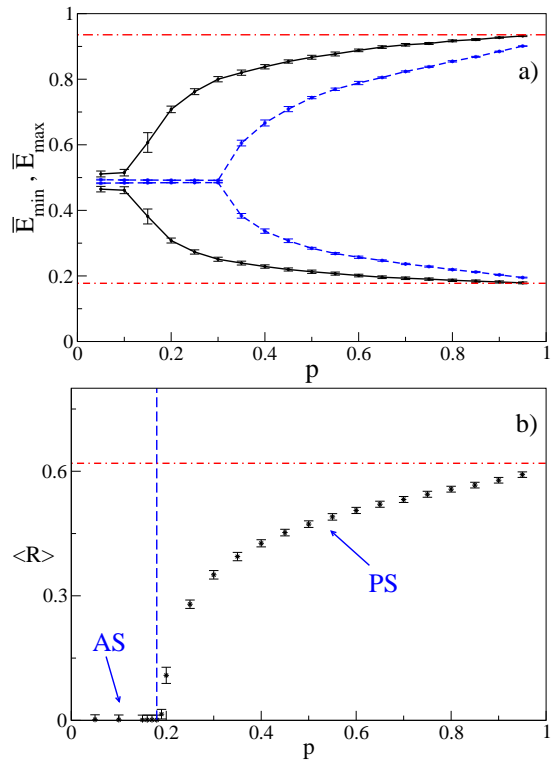

\includegraphics[width=0.4\textwidth]{fig2a}
\includegraphics[width=0.4\textwidth]{fig2b}
\caption{(Color online) a) Minima and maxima of the average field $\bar E$ 
as a function of $p$ for $N=1,600$, the circles joined
by solid lines refer to the annealed disorder, while the stars
connected by the dashed line to the quenched case.
The dot-dashed (red) lines indicate the fully coupled results
(corresponding to $p\equiv 1$). b) Synchronization
indicator $\langle R \rangle$ averaged over time as a function of $p$ for
the quenched case with $N=3,200$. The data refer to ER networks
with constant probability (CP) and 
have been estimated,
after discarding a transient of $4 \times 10^7$ spikes, by
averaging over a train of $1 - 2 \times 10^7$  spikes.
}
\label{fig:2}   
\end{figure}

As an initial reference study we consider ER networks with 
unidirectional links chosen at random with a constant 
probability (CP) $p$ for any network size $N$.
This amounts to consider the limiting case $\gamma \to 1$ 
and to have an average connectivity scaling
linearly with the size, i.e. $\langle k \rangle=p N$.
Let us firstly examine the minima and maxima of $\bar E$ of a network of
size $N$ by varying the probability $p$ between 0 and 1, as shown in 
Fig. \ref{fig:2}a for $N=1,600$. At low probability one has AS while 
the PS regime emerges only for sufficiently large $p$, both in the quenched
and annealed case. This result is related to the fact that the presence of 
noise reduces the coherence needed to observe the PS regime ~\cite{mohanty}.
As a matter of fact increasing $p$ (i.e. diminishing the number of
broken links in the network) the attractor size increases and finally reaches
the fully coupled result. The degree of coherence can be measured
in terms of the average synchronization indicator $\langle R \rangle$. As shown in
Fig. \ref{fig:2}b the system coherence steadily increases with
$p$, except in the AS regime where $\langle R \rangle \sim 0$, apart for 
finite size fluctuations.

\begin{figure}
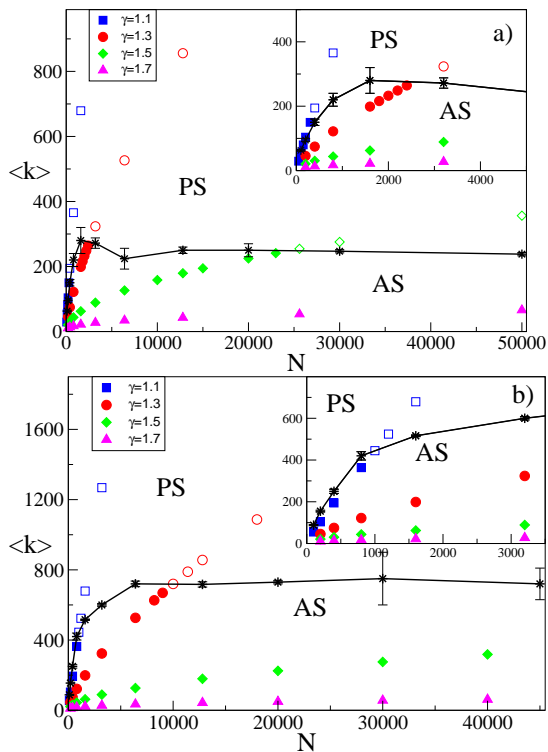

\includegraphics[width=0.4\textwidth]{fig3a.eps}
\includegraphics[width=0.4\textwidth]{fig3b.eps}
\caption{(Color online) Phase diagram for the macroscopic activity of the network
in the $(N,\langle k \rangle)$ plane: a) annealed disorder and b) quenched disorder.
The (black) asteriskes connected by the solid (black) line correspond to 
the transition values $\langle k \rangle_c$ from AS to PS regime estimated for ER networks
with CP. The other symbols refer to ER with $\gamma > 1$: solid (resp. empty) symbols
individuate AS (resp. PS) states. In particular, (blue) squares refer to $\gamma=1.1$,
(red) circles to $\gamma=1.3$, (green) diamonds to $\gamma=1.5$, and (magenta) triangles
to $\gamma=1.7$. The reported data are relative to the state of the network
after discarding transients ranging from $2 \times 10^7$ spikes at the smaller sizes
to $3 \times 10^8$ spikes for the larger networks.
}
\label{fig:3}   
\end{figure}

Let us now report the phase diagram for the macroscopic activity
of the network in the $(N,\langle k \rangle)$ plane, for both annealed and quenched disorder.
Increasing the average connectivity, keeping the system size fixed, leads to a 
transition from AS to PS regimes (see Fig. \ref{fig:3}).
The transition occurs at a critical average connectivity $\langle k \rangle_c$, 
indicates by the asteriskes connected by a black solid line in Fig. \ref{fig:3},
which for low $N$ increases steadily with $N$, 
but eventually saturates for $ N  >  10,000$
to an asymptotically constant value which depends on the noise realization: namely, 
$\langle k \rangle_{as}=725\pm25$ for the quenched case and $\langle k \rangle_{as}=225\pm25$ in the annealed one
\cite{note3}.

\subsection{ER networks with $\gamma$-dependent probability} 

\begin{figure}
\includegraphics[width=0.4\textwidth]{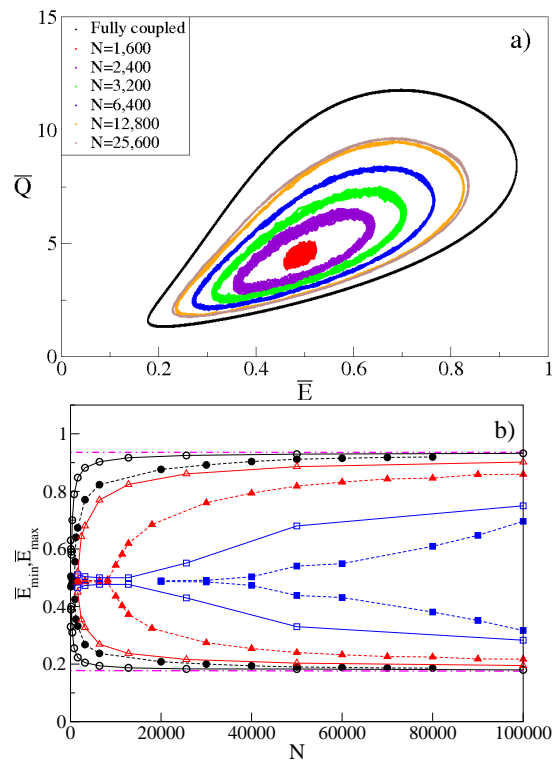}
\includegraphics[width=0.4\textwidth]{fig4b.eps}
\caption{(Color online) a) Macroscopic attractors in the $(\bar E, \bar Q)$ plane for a
ER networks with $\gamma=1.3$ and annealed disorder,
the curves from the interior to the exterior corresponds
to increasing system sizes, from $N=1,600$ to 25,600.
The most external (black) curve refers to a fully coupled
network with $N=3,200$.
b) Minima and maxima values of the average field $\bar E$ 
as a function of $N$ for various $\gamma$-values: namely,
(black) circles $\gamma=1.1$, (red) triangles $\gamma=1.3$
and (blue) squares $\gamma=1.5$. The empty (resp. filled)
symbols refer to annealed (resp. quenched) disorder.
The dot-dashed (magenta) lines
indicate the fully coupled values. 
}
\label{fig:3bis}   
\end{figure}

To verify the generality of these results we investigate
ER networks with $\gamma$-dependent probability. In particular, 
we have estimated, for system sizes in the range $100 < N < 200,000$,
the macroscopic attractors for various $\gamma$-values 
(namely, $\gamma=1.1$, 1.3, 1.5 and 1.7), after discarding long
transient periods. For small system sizes the network is in the AS
regime which is characterized by a spot-like attractor in the $(\bar E,\bar Q)$-plane.
For larger number of neurons, PS emerges in the system characterized by
closed curve (macroscopic) attractors.
Furthermore, similarly to the results reported in Ref. \cite{olmi}, increasing $N$ 
the curves tend to  an asymptotic shape, corresponding to the fully coupled attractor, while 
fluctuations diminish. To exemplify this point various macroscopic attractors for $\gamma=1.3$ and annealed
disorder are reported in Fig. \ref{fig:3bis}a. 

As already reported for the CP networks the systems with annealed disorder
converge more rapidly with $N$ towards the asymptotic fully coupled attractor 
with respect to the quenched case, as shown in Fig. \ref{fig:3bis}b for $\gamma=1.1$, 1.3, 1.5.
Increasing $\gamma$ we observe that the transition from AS to PS
occurs at larger and larger system size, both for annealed and quenched
disorder. The results for $\gamma =1.7$ are not shown in Fig. \ref{fig:3bis}b,
since for all the examined network sizes (up to $N=200,000$) $\bar E$ extrema 
coincide within the error bar, indicating that the system is in the AS regime
\cite{note4}.

AS (resp. PS) regimes are reported in the
phase diagram displayed in Fig. \ref{fig:3} 
as filled (resp. empty) symbols 
for the investigated $\gamma$-values. 
The critical line $\langle k \rangle_{c} = \langle k \rangle_{c}(N)$
(indicated by the solid black line in Fig. \ref{fig:3})
denoting the transition from AS to PS coincides with that
determined previously for CP networks.
We can thus safely affirm that the dynamical regimes of the
ER networks depend, at a macroscopic level, simply on the average connectivity,
once the system size $N$ is fixed. This could be expected from the fact that
for ER networks the distribution of links per neuron is completely determined
by $\langle k \rangle$ (see Eq. (\ref{poisson})). However, the independence of 
$\langle k \rangle_c$ from $N$ at $N > 10,000$ is unexpected and suggests that coherent behaviors, 
like PS regimes, can be observed in networks of any size for vanishingly
small relative connectivity $\langle k \rangle/N$.

\section{Network Homogeneity}
\label{sec:3}

We observe that for any considered $\gamma$ the fields $(E_i,Q_i)$ associated 
with the different neurons tend to synchronize for increasing
$N$. Therefore in the thermodynamic limit ($N \to \infty$) 
disordered networks tend to behave as fully coupled
ones, where all the neurons are equivalent and a single field is
sufficient to describe the macroscopic evolution of the system.

\begin{figure}
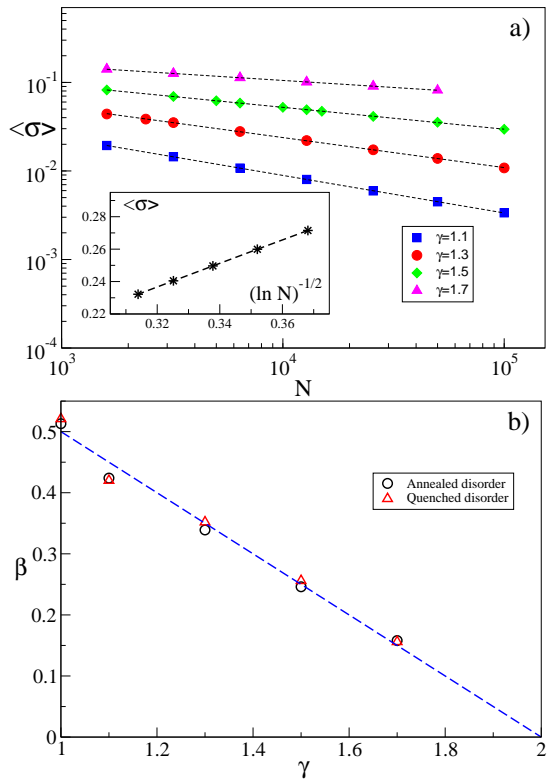

\includegraphics[width=0.4\textwidth]{fig5a.eps}
\includegraphics[width=0.4\textwidth]{fig5b.eps}
\caption{(Color online) a) Average standard deviation $\langle \sigma \rangle$ versus
the system size $N$ for annealed disorder and 
various $\gamma$-values: $\gamma=1.1$
(blue) squares, $1.3$ (red) circles, $1.5$ (green) diamonds and
$1.7$ (magenta) triangles. The dashed line represents best fits with a 
power-law $N^{-\beta}$ to the reported data. The data in the inset
(black asterisks) refers to a $\gamma=2.0$, the dashed line is a guide for the eyes.
b) Power-law exponents $\beta$
for annealed (black circles) and quenched (red triangles) disorder in the network
as a function of the parameter $\gamma$. The dashed (blue) line refers
to the linear law $\beta = 1 - \gamma/2$.
The reported data have been estimated by averaging over
trains made of $2 \times 10^6 - 10^8$  spikes, after
discarding transients of $4 \times 10^5 - 4 \times 10^6 $ spikes.
}
\label{fig:4}       
\end{figure}

In order to quantify the level of homogeneity among the various neurons,
we measured the standard deviation (Eq. (\ref{eq:sigma})) relative 
to the fluctuations of the different local fields
$E_i$  with respect to the average field $\bar E$.
In Fig. \ref{fig:4}a we plot the time average of the standard deviation, 
$\langle \sigma \rangle$, for annealed
disorder and various $\gamma$ values. 
We observe a power law decay $\langle \sigma \rangle \propto N^{-\beta}$, where
the exponent $\beta$ depends on the $\gamma$-parameter as
$\beta = 1 - \gamma/2$ (see Fig.~\ref{fig:4}b).
Furthermore, for the liming case $\gamma=2.0$ 
the decay of $\langle \sigma \rangle$ is consistent with
a scaling $1/\sqrt{\ln N}$ as displayed in the inset of Fig.~\ref{fig:4}a.
Alltogether, the reported dependencies suggest the following relationship to hold:
\begin{equation}
\langle \sigma \rangle \propto \frac{1}{\sqrt{\langle k \rangle}}
\quad ;
\label{sigma_scal}
\end{equation}
thus fields fluctuations are driven
by the average in-degree value irrespectively of the total number of neurons.
These results confirm once more that in the limit $N \to \infty$
the neural field dynamics converges to that of homogeneous networks,
for both quenched and annealed disorder. 

The relationship among $\langle \sigma \rangle$ and the
average connectivity reported in eq. (\ref{sigma_scal}) clearly
indicates that for a sparse network, with constant
average connectivity, $\langle \sigma \rangle$ will remain finite 
even in the thermodynamic limit.
Therefore, we can conclude that a sparse network cannot be ever reduced 
to a fully coupled one by simply rescaling the synaptic coupling as done 
in Ref. \cite{olmi}, even for very large systems.

\section{Chaotic vs Regular Dynamics} 
\label{sec:4}

Homogeneous fully connected pulse coupled networks 
exhibit regular dynamics \cite{vvres}. In particular,
for excitatory network and finite pulses the AS becomes 
a {\it splay state} characterized by all neurons spiking one
after the other at regular intervals with the same frequency 
and by a constant mean field $\bar E$, while the PS regime becomes
perfectly periodic at a macroscopic level~\cite{abbott}.

The introduction of disorder in the network leads to irregularity
in the dynamics of the single neurons, which are reflected also
at a macroscopic level. This kind
of deterministic irregular behavior has been identified
as {\it weak chaos} whenever the irregularity vanishes for sufficiently 
large system size~\cite{popo}. The chaotic motion can be characterized in terms
of the maximal Lyapunov exponent $\lambda$: regular orbits have
non positive exponents, while chaotic dynamics are associated 
with $\lambda  >  0$.

\begin{figure}
\includegraphics[width=0.4\textwidth]{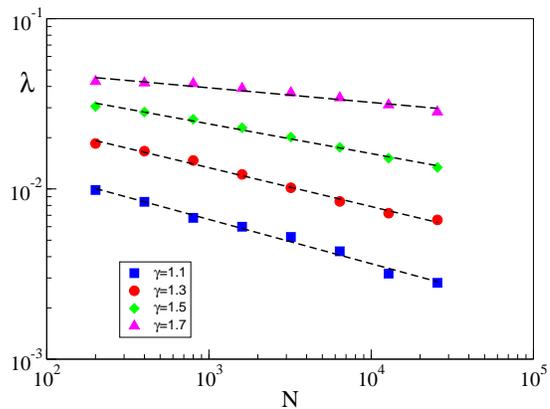}
\caption{(Color online) Maximal Lyapunov exponents as a function of
the system syze $N$ for various $\gamma$-values.
The data have been obtained by discarding a transient
of the order of $10^8 -10^9$ spikes 
and then by following the dynamics in the real and
tangent space for an equivalent duration,
moreover the data have been averaged over 
3 to 5 different network realization with quenched disorder.
}
\label{fig:5}       
\end{figure}

For finite size networks chaotic dynamics is observed both for annealed 
and quenched disorder.  However, for all 
$\gamma$-values examined in this work, $\lambda$ tends to decrease for
a sufficiently large number of neurons in the network.
Therefore we can safely affirm that for
any ER network with average connectivity given by Eq.
(\ref{kave}) the neuronal dynamics is weakly chaotic; i.e.
the evolution will become completely regular for infinite networks.
Numerical results for quenched disorder are reported in
Fig. \ref{fig:5}a for various $\gamma$-values.
The maximal Lyapunov exponent exhibits clear power-law 
decays $N^{-\delta}$, with $\delta$ decreasing
and eventually vanishing for $\gamma \to 2$ (see Table \ref{tab1}). 
These results generalize previous indications reported in
Olmi et al. for a specific realization of diluted network~\cite{olmi}.

%
%
\begin{table}
\caption{Power law exponents giving the decay of the maximal
Lyapunov $\lambda \propto N^{-\delta}$ for the data reported
in Fig. \ref{fig:5}a for quenched disorder.}
\label{tab:1}       
\begin{tabular}{ll}
\hline\noalign{\smallskip}
$\gamma$ &  \hfill  $\delta$ \hfill \\
\noalign{\smallskip}\hline\noalign{\smallskip}
1.1 & $0.26 \pm 0.01$ \\
1.3 & $0.228 \pm 0.007$ \\
1.5 & $0.174 \pm 0.006$ \\
1.7 & $0.085 \pm 0.009$ \\
\noalign{\smallskip}\hline
\label{tab1}       
\end{tabular}
\end{table}

Considering networks with annealed disorder it must be underlined that
for sufficiently large number of neurons and $\gamma <  2$,
the maximal Lyapunov reveals a tendency to decrease. However,
clear scaling laws cannot be inferred for all the examined $\gamma$-values
on the range of affordable system sizes, namely $50 \le N \le 12,500$.
Larger network sizes are probably required with unreliable synapses
to have clear scaling laws, due to the fact that the transition
from AS to PS occurs at critical system sizes within the investigated range.
On the contrary, in the quenched case once $\gamma$ is fixed
an unique regime is observable for almost all the investigated
sizes. In particular, the network is always in an AS for
$\gamma=1.7$ and 1.5, while it is in the PS
regime for $\gamma=1.1$ and 1.3 for (almost) all the considered
number of neurons. These additional findings strengthen the above 
reported conclusions: finite size systems are weakly chaotic 
for any considered dynamical regime.

According to the results reported above, we expect that for sparse
networks the maximal Lyapunov exponent will eventually saturate to 
some constant value, apart for possible logarithmic corrections.
Thus, sparse networks should remain chaotic for 
any system size, paralleling the behavior of the
microscopic fluctuations shown in the previous Section.
Therefore microscopic inhomogeneities and chaotic behavior
appear as deeply related.

\section{Discussion} 
\label{sec:5}

Collective periodic oscillations in excitatory ER networks
can be observed only above a critical average in-degree.
This latter quantity saturates to a constant value for 
networks with a sufficiently large number of neurons, 
thus suggesting that the key ingredient responsible for
the emergence of collective behaviors is the number of 
pre-synaptic neurons both for massively connected networks
as well as for sparsely connected ones.
This result confirms and generalizes previous 
findings on the stability of the complete synchronized state for 
pulse-coupled Hindmarsh-Rose neurons \cite{belykh}. 
Furthermore, our results indicate that the minimal network size 
required to observe collective oscillations diverges
with the exponent $\gamma$ ruling the scaling of the
average connectivity with the number of neurons.

The presence of annealed disorder in the network 
(corresponding to unreliable synapses)
favors the emergence of coherent activity with respect to
the quenched case (associated with reliable synapses), since 
in the first case the asymptotic average connectivity required 
to observe collective oscillations is much smaller.
This is probably due to the fact that in the annealed 
situation each neuron is on average subjected to the same 
train of stimuli, while with quenched disorder 
the dynamics of each neuron depends heavily on its neighbors.
Furthermore, for sufficiently large networks the macroscopic 
behavior observable with reliable or unreliable synapses 
becomes identical. This seems to indicate that
at the level of population dynamics the reliability or 
unreliability in the synaptic transmission can be irrelevant.

The average in-degree $\langle k \rangle$ also controls the 
fluctuations among different neurons, being the fluctuations 
proportional to the inverse of the square root of $\langle k \rangle$.
Therefore, for ER networks with average in-degree proportional 
to any positive power of $N$, the fluctuations will vanish in the
limit $N \to \infty$, leading to a homogeneous 
collective behavior analogous to that of fully connected
networks. On the contrary, inhomogeneities among neurons 
will persist at any system size in sparse networks.

Recent experimental results on the intact neuronal network of
the barrel cortex of anesthetized rats seems to clearly
suggest that the dynamics of this system is chaotic \cite{london}.
This result poses severe questions about the possibility of 
reliable neural coding. The evolution of our models is chaotic for
any finite networks. However, in presence of coherent
periodic activity, the chaoticity present in the
system is not so strong to destroy the average collective motion. 
Thus the trial-to-trial variability induced by chaos
does not prevent the possibility of the network to encode 
information. The information can be coded in some 
property associated to the global
oscillations of the network which are robust
to local fluctuations, and this {\it collective coding}
can represent an alternative to the dilemma rate versus spike 
timing coding~\cite{london}.

The level of chaos in the examined networks 
decreases with $N$ and the dynamics becomes
regular in the thermodynamic limit.
However, the decrease of the maximal
Lyapunov exponent with $N$ slows down dramatically 
for $\gamma \to 2$, suggesting an erratic
asymptotic behavior for sparse networks.

Our results represent only a first step in the
analysis of what matters in the network topology
for the emergence of coherent neural dynamics.
In future works, our findings should be critically verified for 
more complex topologies, like scale-free and small-world,
and the influence of other ingredients, like
the asymmetry in the in-degree and out-degree distributions
\cite{roxin}, should be also addressed.

\begin{acknowledgements}
We acknowledge useful discussions with A. Politi, A. Pikovsky,
P. Bonifazi and M. Timme. This research project is part of the activity
of the Joint Italian-Israeli Laboratory on Neuroscience
funded by the Italian Ministry of Foreign Affairs and it
has been partially realized thanks to the support of CINECA
through the Italian Super Computing Resource Allocation
(ISCRA) programme, project ECOSFNN. One of us (LT) thanks
HPC-Europa2 programme for supporting him during the period
spent in the Network Dynamics Group of the Max Planck Institute
for Dynamics and Self Organization in G\"ottingen (Germany).
\end{acknowledgements}


\begin{thebibliography}{}
\bibitem{note1} 
A sparse network is usually defined as a network where
the average connectivity remains constant in the thermodynamic limit, see \cite{golomb}.
\bibitem{note2} 
In the original definition reported in \cite{winfree},
$T_{q,n}=T_{j,n}$, thus ensuring that the phase is bounded between 0 and $2\pi$.
However this definition requires to memorize all the spiking events within a certain
time interval and then estimate the phases and $R(t)$ over such time
span. Our definition does not guaratee that the phase is strictly bounded within the $[0:2\pi]$-interval, 
but it allows to estimate $R(t)$ during the simulation run without memorizing any event.
Indeed we have verified that, at least in the studied cases, no
peculiar differences are observable by employing the two phase definitions
and this is probably due to the fact that the dynamics of all the neurons in the
networks are statistically equivalent.
\bibitem{note3} 
We have verified that the above reported scenario is not modified by considering fully 
homogeneous connectivity with $P(k)=\delta(k-k_0)$, the only difference concerns the asymptotic
value  $\langle k \rangle_{as}$ at which the transition from AS to PS occurs. As a matter
of fact $\langle k \rangle_{as}$ is slightly smaller than for ER distributions due to
the absence of fluctuations among the in-degree values in the homogeneous case.
\bibitem{note4} 
A simple estimation of the minimal network size needed to observe 
PS for generic $\gamma$-distributions can be obtained by inverting 
Eq. (\ref{kave}) for the asymptotic values $\langle k \rangle_{as}$, namely
\begin{equation}
N_c = \left[  \frac{p}{2-\gamma} \langle k \rangle_{as} + 1 \right]^{1/(2-\gamma)}
\quad ;
\label{ncrit}
\end{equation}
for $\gamma=1.7$ this leads to $N_c = 131,784,000$ (resp. $N_c = 2,740,117$)
in the quenched (resp. annealed) case. This confirms that coherent
activity is unobservable for $\gamma=1.7$ with our computational resources
for the chosen coupling parameter, namely $g=0.4$.
\end{thebibliography}


\end{document}